%% file: AP_Paper_arxiv.tex
\pdfoutput=1

\documentclass[10pt,a4paper,twoside]{article}
\usepackage[english]{babel}
\usepackage[pdftex]{graphicx}

\usepackage[T1]{fontenc}
\usepackage[latin1]{inputenc}
\usepackage{amsmath}
\usepackage{caption}

\usepackage{bibgerm}

\makeatletter
\renewcommand{\section}{\@startsection{section}{2}{0cm}{-\baselineskip}
{0,5\baselineskip}{\normalsize\bfseries}}
\renewcommand{\subsection}{\@startsection{subsection}{3}{0cm}{-\baselineskip}
{0,5\baselineskip}{\normalsize\slshape}}
\makeatother

\usepackage{subfig}
\usepackage{array}
\usepackage{booktabs}
\usepackage{upgreek}
\usepackage{amstext}

\newcommand{\SI}[2]{\ensuremath{#1 \, {\rm #2}}}

\DeclareMathAlphabet{\mathcal}{OMS}{cmsy}{b}{n}
\newcommand{\eps}{\text{\large\usefont{OML}{cmr}{m}{n}\symbol{15}}}
\newcommand{\rrho}{\text{\large\usefont{OML}{cmr}{m}{n}\symbol{37}}}

\DeclareMathVersion{mathchartertext}
\SetSymbolFont{letters}{mathchartertext}{OML}{mdbch}{m}{n}

\begin{document}

\input{sections/title.tex}

\input{sections/section_1.tex}

\input{sections/section_2.tex}

\input{sections/section_3.tex}

\input{sections/section_3_1.tex}

\input{sections/section_3_2.tex}

\input{sections/section_3_2_1.tex}

\input{sections/section_3_2_2.tex}

\input{sections/section_3_2_3.tex}

\input{sections/section_4.tex}

\input{sections/section_4_1.tex}

\input{sections/section_4_2.tex}

\input{sections/section_5.tex}

\input{sections/section_6.tex}

\input{sections/acknowledgments}

\end{document}

%% file: sections/title.tex
\title{Afterpulse Measurements of R7081 Photomultipliers for the Double Chooz Experiment}
\author{Julia Haser$^{b*}$, Florian Kaether$^b$, Conradin Langbrandtner$^b$, \\ Manfred Lindner$^b$, Sebastian Lucht$^a$, Stefan Roth$^a$, Manuel Schumann$^a$, Achim Stahl$^{a}$, \\ Anselm St\"{u}ken$^{a}$\thanks{Corresponding authors.}~ and Christopher Wiebusch$^a$ }
\date{
 \small \sl 
{$^a$}RWTH Aachen, III. Physikalisches Institut B, \\
 Otto-Blumenthal-Stra\ss{}e, D-52056 Aachen, Germany \\
 E-mail: {\tt anselm.stueken@physik.rwth-aachen.de} \\
 \vspace{0.2cm}
{$^b$}Max-Planck-Institut f\"{u}r Kernphysik, \\ 
 Saupfercheckweg 1, D-69117 Heidelberg Germany. \\
 E-mail: {\tt Julia.Haser@mpi-hd.mpg.de} \\ 
\vspace{0.4cm}
 {\small (Submitted to Journal of Instruments (JINST))}
}

\maketitle

\begin{abstract}
We present the results of afterpulse measurements performed as qualification test for 473 inner
detector photomultipliers of the Double Chooz experiment. The measurements include the determination of a total
afterpulse occurrence probability as well as an average time distribution of these pulses. Additionally, more
detailed measurements with different light sources and simultaneous charge and timing measurements were performed with a
few photomultipliers to allow a more detailed understanding of the effect. The results of all measurements are presented and discussed.
\end{abstract}

%\keywords{Photomultiplier; Afterpulse; Double Chooz}

%% file: sections/section_1.tex
\section{Introduction}
The reactor antineutrino disappearance experiment Double Chooz is designed to determine the leptonic mixing angle $\theta_{13}$ \cite{DCProposal}.
A disappearance of reactor $\overline{\nu}_e$ with a best fit value of
\begin{equation}
{\rm sin}^2 2 \theta_{13} = \, 0.109 \, \pm \, 0.030 \, ({\rm stat}) \, \pm \, 0.025 \, ({\rm sys})
\end{equation}
was already presented in \cite{DC1stPub, DC2ndPub}.\\
The experimental design consists of two identical detectors with a \SI{10.3}{m^3} target each. 
The target material is a gadolinium doped liquid scintillator in which neutrinos are detected by inverse beta decay. 
To detect the scintillator light each target is surrounded by 390 photomultiplier tubes (PMT), 
which are 10 inch, hemispherical PMTs from Hamamatsu (type R7081MOD-ASSY) with a bialcali photokathode (Sb-K-Cs). 
For the two detectors a total of 780 PMTs + 23 spares were provided, about one half by German and the other half by Japanese groups.\\
For a preselection and validation of the specifications of the 803 target PMTs two test setups were built: (A) at the ``Max-Planck-Institut f\"{u}r
Kernphysik'' (MPIK) in Heidelberg in collaboration with the ``RWTH Aachen University'' \cite{Seb} and (B) at the
``Tokyo Institute of Technology'' \cite{JapanPMTpaper}. This article focuses on afterpulse measurements that have been performed with setup A.\\
For 473 PMTs the probability of the occurrence of afterpulses has been measured.
Furthermore the average arrival time distribution of the afterpulses and their amplitudes were determined. 
Two different analysis methods of the same data set will be presented in this article. 
Additionally, more detailed investigations have been done including a combination of correlated charge and timing measurements of the afterpulses. 
Different light sources and hardware components were used to validate the consistency of the different measurements. 
The more detailed measurements were done only for a few selected PMTs.

%% file: sections/section_2.tex
\section{Afterpulse probability} \label{sec:AfterpulseProb}
Photoelectrons created at the photocathode of a photomultiplier (PMT) may cause ionization of the residual gas between the photocathode and the first dynode.
Due to the electric field the positive ions travel back to the photocathode and create secondary electrons, which result in afterpulses.
The photoelectrons of the afterpulses may themselves create afterpulses, which are called afterpulses of higher order.\\
In this article the expected number of afterpulses $\mathcal{E}$ induced by a single photoelectron (SPE) is determined, including also higher order afterpulses.
Furthermore, the afterpulse probability $\mathcal{P}$, considering only first order afterpulses, is calculated. 
We define $\mathcal{P}$ to be the probability of measuring at least one afterpulse induced by an initial SPE. 
The relation between $\mathcal{E}$ and $\mathcal{P}$ will be discussed at the end of this section.\\
\newline
Multiple ionizations induced by one single photoelectron are treated as independent processes.
This leads to the conclusion that the probability $P_{\mu}(n,k)$ of creating $k$ first order afterpulses induced by $n$ initial photoelectrons obeys
Poissonian statistics:
\begin{equation}
\label{eq:APProb_nMu}
P_{\mu}(n,k) = \frac{(n \cdot \mu)^k}{k!} {\rm e}^{- n \cdot \mu} \, .
\end{equation}
Given that each ionization process results in an afterpulse, $\mu$ represents the average number of ionization processes per photoelectron.
Based on this equation the probability $P_{\mathrm{ap}}(n)$ of measuring at least one afterpulse for $n$ initial photoelectrons is given by
\begin{equation}
\label{eq:APProb}
P_{\mathrm{ap}}(n) = P_{\mu}(n,k \ge 1) = 1 - {\rm e}^{-n \cdot \mu} \, .
\end{equation}
We define $\mathcal{P}:=P_{\mathrm{ap}}(1)$, so equation \ref{eq:APProb} can be expressed as
\begin{equation}
\label{eq:CorrPap(n)}
P_{\mathrm{ap}}(n) = 1- (1-\mathcal{P})^n .
\end{equation}
To calculate the expected number of afterpulses we start with the expected value of first order afterpulses $E_{\mathrm{1st}}(n)$ induced by $n$ initial
photoelectrons. Using equation \ref{eq:APProb_nMu} we find:
\begin{equation}
\label{eq:E1st(n)}
E_{\mathrm{1st}}(n) = \sum_{k=0}^{\infty} P_{\mu}(n,k) \cdot k = n \cdot \mu \, .
\end{equation}
In order to include second order afterpulses it is necessary to consider that afterpulses can contain more than one photoelectron.
According to our model, the afterpulse charge $\ell$ is thought to be independent of the number of initial photoelectrons $n$ and the afterpulse order.
We therefore introduce the expected number of photoelectrons contained by an afterpulse
$\overline{\ell} = \sum_{\ell=0}^{\infty} (\ell \cdot \rrho_{\ell})$, while $\rrho_{\ell}$ represents the time-averaged afterpulse charge density histogram.
% we introduce the time-averaged afterpulse charge density histogram $\rrho_{\ell}$.
% With $\overline{\ell} = \sum_{\ell=0}^{\infty} (\ell \cdot \rrho_{\ell})$ as expected number of photoelectrons contained by an afterpulse
The expected number of second order afterpulses can now be written as
\begin{eqnarray}
E_{\mathrm{2nd}}(n) & = & \sum_{k=0}^{\infty} k \cdot P_{\mu}(n,k) \cdot \sum_{k'=0}^{\infty} \sum_{\ell=0}^{\infty} k' \cdot P_{\mu}(\ell,k') \cdot \rrho_{\ell} \nonumber \\  
{} & = & \sum_{\ell=0}^{\infty} n \mu \cdot \ell \mu \cdot \rrho_{\ell} = n \mu^2 \cdot \overline{\ell} \,.
\end{eqnarray}
For the number of expected afterpulses $E_i(n)$ of the i-th afterpulse order we can then conclude:
\begin{equation}
\label{eq:Eith(n)}
E_i(n) = E_{i-1}(n) \cdot \mu \cdot \overline{\ell} = n \mu \cdot (\mu \cdot \overline{\ell})^{i-1} \, .
\end{equation}
Summation of all orders leads to a total number of expected afterpulses of
\begin{equation}
\label{eq:Eap(n)}
E(n) = n \cdot \mu \cdot \sum_{i=1}^{\infty} ( \mu \cdot \overline{\ell} )^{i-1} = \frac{n \cdot \mu}{1-\mu \cdot \overline{\ell}} = n \cdot \mathcal{E} \, ,
\end{equation}
while $\mathcal{E}:=E(1)$ is defined as expected number of afterpulses induced by a single initial photoelectron.
Combining equation \ref{eq:APProb} and \ref{eq:Eap(n)} $E(n)$ can be expressed as a function of $P_{\mathrm{ap}}(n)$:
\begin{equation}
\label{eq:Eap(Pap)}
E(n) = -\frac{n \cdot {\rm ln}(1-P_{\mathrm{ap}}(n))}{n+{\rm ln}(1-P_{\mathrm{ap}}(n)) \cdot \overline{\ell}} \, .
\end{equation} 
In our measurement we will record the occurrence of afterpulses in the time range up to \SI{16}{\upmu s}, as the selection specification of the
afterpulse probability was defined for this interval.
In section \ref{subsec:comp} we will show that first order afterpulses occur only in this time range and thus the measurement of $\mu$ will not
be affected by the measurement's constraint.
However, afterpulses of higher order may occur later than \SI{16}{\upmu s} after the initial pulse and will not be detected in the measurement.
In order to account for this effect, we separate equation \ref{eq:Eap(n)} in first order and higher order afterpulses $E_{\mathrm{h}}$:
\begin{eqnarray}
\label{eq:Eap1st+Eapho}
\mathcal{E} & = & \frac{\mu}{1-\mu \cdot \overline{\ell}} = \mu + E_{\mathrm{h}} \, ,\\
\label{eq:Eapho}
E_{\mathrm{h}} & = & \frac{\mu^2 \cdot \overline{\ell}}{1-\mu \cdot \overline{\ell}} =  \mu \cdot \overline{\ell} \cdot \mathcal{E} \, .
\end{eqnarray}
We define $\mathcal{E}^{\Delta}$ and $E^{\Delta}_{\mathrm{h}}$ to be the expected number of afterpulses within the time range from 0 to \SI{16}{\upmu s}.
Since only a fraction of higher order afterpulses
\begin{equation}
\label{eq:f}
f = \frac{E^{\Delta}_{\mathrm{h}}}{E_{\mathrm{h}}} \, ,
\end{equation}
will be detected, we redefine equation \ref{eq:Eap1st+Eapho}:
\begin{equation}
\label{eq:Eap0-16}
\mathcal{E}^{\Delta} = \mu + E_{\mathrm{h}} \cdot f = \frac{ \mu - \mu^2 \cdot \overline{\ell} \cdot (1-f)}{1-\mu \cdot \overline{\ell}} \,.
\end{equation}
In our analyses we will determine the discrete temporal distributions $\mu_i$ and $\mathcal{E}_i$ for the expected number of first and first plus
higher order afterpulses in a time bin $i$. 
Based on equation \ref{eq:Eapho} the temporal distribution of the expected number of higher order afterpulses $E_{\mathrm{h},i}$ can be calculated via
\begin{equation}
\label{eq:Eho(mu,ti)}
E_{\mathrm{h},i} = \overline{\ell} \cdot \sum_{k=0}^{i} \mathcal{E}_{k} \cdot \mu_{i-k} \, .
\end{equation}
The fraction $f$ is then computed via the expression
\begin{equation} 
\label{eq:fint}
f = \frac{E^{\Delta}_{\mathrm{h}}}{E_{\mathrm{h}}} = \frac{\sum_{i}^{\SI{16}{\upmu s}} E_{\mathrm{h},i}}{\sum_{i}^{\infty} E_{\mathrm{h},i}} \, ,
\end{equation} 
which is independent of $\overline{\ell}$.

%% file: sections/section_3.tex
\section{Timing measurements with 473 PMTs} \label{sec:TimingMeasurement}
Timing measurements of 473 PMTs were performed to evaluate the time-dependent occurrence of afterpulses.
Additionally, the total afterpulse occurrence for the time range from 0 to \SI{16}{\upmu s} after an initial signal was
determined to validate the PMT specification for the Double Chooz experiment.
The selection specification demaned a total afterpulse probability of less than \SI{10}{\%}.\\
In this section we will present two methods to analyse the same afterpulse data set. 
At first the expected value of afterpulses $\mathcal{E}$ produced per SPE event is calculated. The second analysis method
will bring forth the expected value of first order afterpulses $\mu$, from which the afterpulse probability $\mathcal{P}$ can be derived.
Moreover, the average number of photoelectrons $\overline{\ell}$ per afterpulse can be determined using
$\mathcal{E}$ and $\mathcal{P}$ (see section \ref{subsec:comp}).

%% file: sections/section_3_1.tex
\subsection{Experimental setup}
\begin{figure}
  \centering
  \includegraphics[angle=270,width=0.9\textwidth]{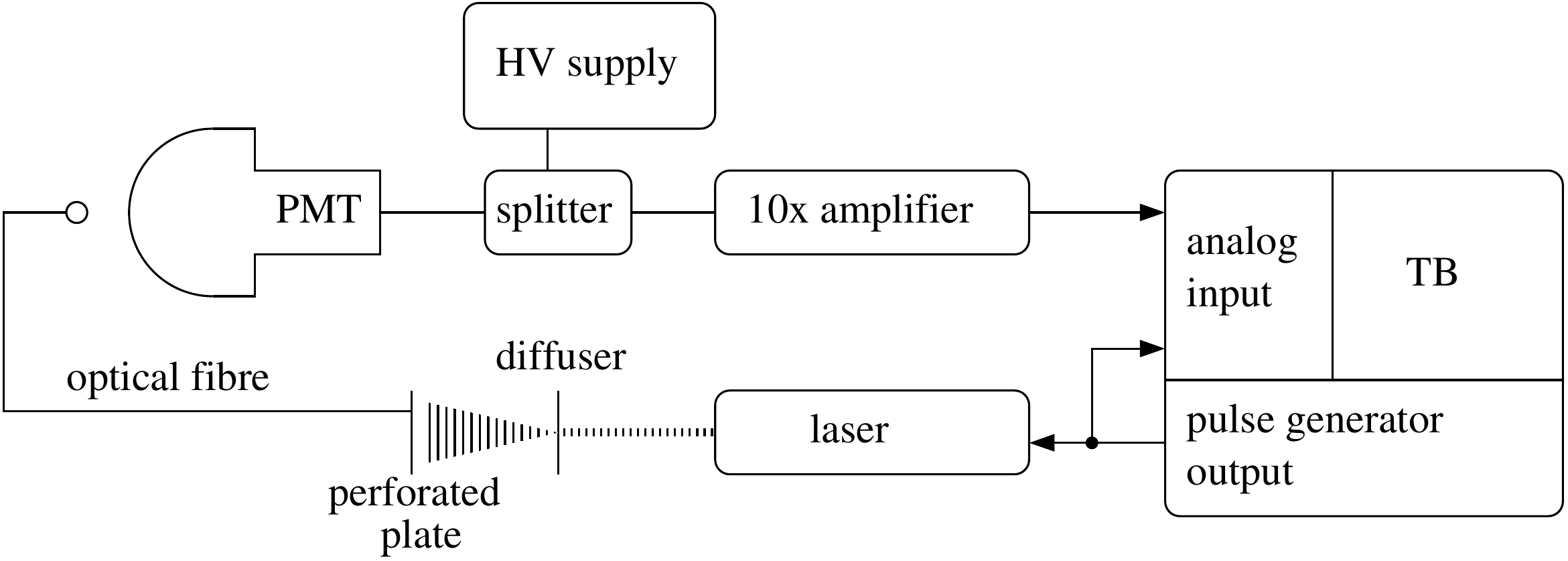}
  \caption{Scheme of the experimental setup.}
  \label{fig:scheme_setup}
\end{figure}

The experimental setup of the timing measurements is sketched in figure \ref{fig:scheme_setup}. 
It is a fraction of the complete experimental setup which was build at MPIK to allow for the qualification tests of 473 PMTs. 
For more detailed information about the complete setup and all performed measurements the reader may refer to \cite{Seb}.\\
The PMTs were mounted on a rack system inside a light tight and electromagnetically shielded Faraday room.
It allowed the calibration of 30 PMTs in parallel. The data acquisition system as well as the power supply was installed outside the Faraday room. 
A splitter box (a module designed and built for the DC experiment by the CIEMAT group, Madrid) was responsible for decoupling the signal from
the supply voltage. Each box was connected to one channel of a high voltage power supply (SY 2527 Universal Multichannel Power Supply System, CAEN)
using a positive polarity to operate the PMTs. 
The signal outputs were connected via a 10x amplifier (sixteen channel amplifier, Model 776, Philips Scientific) to the measurement devices.\\  

For the afterpulse measurement two trigger boards (TB), which were developed for the Double Chooz trigger system \cite{TrigPaper},
have been used. These boards contain 18 analogue input channels with two discriminators each, a sum signal line of all input channels with four
discriminators, a rate counter for each discriminator line, a FIFO memory, which is capable of storing 126 events simultaneously, and a logical
processing unit for the discriminator signals.
The information which is stored in the FIFO for each event contains the rate counters, the discriminator's status of the triggered event as well
as the time difference to the previous event with a resolution of \SI{16}{ns}.
The logical processing unit is freely programmable and based on logical OR and AND operations.
For the measurement only one discriminator for each analogue input channel was used. 
Each channel whose discriminator produces a signal creates an internal event. 
Storing the discriminator status makes it possible to determine the corresponding channel.
For the calibration of 30 PMTs two TBs where used, while each was connected to 15 PMTs. 
The discriminator for each PMT channel was set to \SI{25}{\%} of a single photoelectron pulse (SPE pulse).
The TBs are able to generate a NIM-based logic signal, triggered by an external or internal (via software) signal. 
This signal is used to trigger the light source and its arrival time was measured as strobe in a free channel of each TB.\\
\newline
As light source the beam of a \SI{438}{nm} picosecond injection laser (PiLas) from Advanced Laser Diode Systems was sent through a diffuser to a
perforated plate with 30 optical fibers connected. The other end of the optical fibers were positioned in front of the PMT, illuminating the full
cathode. In order to yield SPE events as signal the laser intensity was adjusted to a level where only about one out of 10 trigger signals led to
a PMT signal. The average signal-to-trigger ratio was $R_{\mathrm{T}}=\SI{9.95}{\%} \pm \SI{2.98}{\%}$, while the error was derived from the variation
in the ratios of all PMTs. It was not possible to adjust the light intensity at each PMT individually, as 30 PMTs were illuminated by the same
laser beam.\\
The applied high voltages had been calibrated previously, so a gain of $10^{7}$ was adjusted at each PMT\cite{Seb}.\\
\newline
During each measurement one TB released a signal to trigger the laser. The strobe signal to each TB defines the start time $t_0$ for one measurement cycle. For all subsequent signals passing the discriminator threshold the time difference to the preceding signal was stored up to a maximum time difference of \SI{20}{\upmu s}. The average time difference between the start time $t_0$ and the detection of the laser signal is \SI{0.7}{\upmu s}. Hence, the collection of all subsequent signals up to \SI{16}{\upmu s} after the laser signal is ensured. For one measurement typically 500\,000 trigger signals were sent, which corresponded to an average number of 50\,000 SPE signals for each PMT.

%% file: sections/section_3_2.tex
\subsection{Analyses and results}\label{subsec:methodresults}
\begin{figure}
  \centering
  \subfloat[]{
    \includegraphics[width=0.47\textwidth]{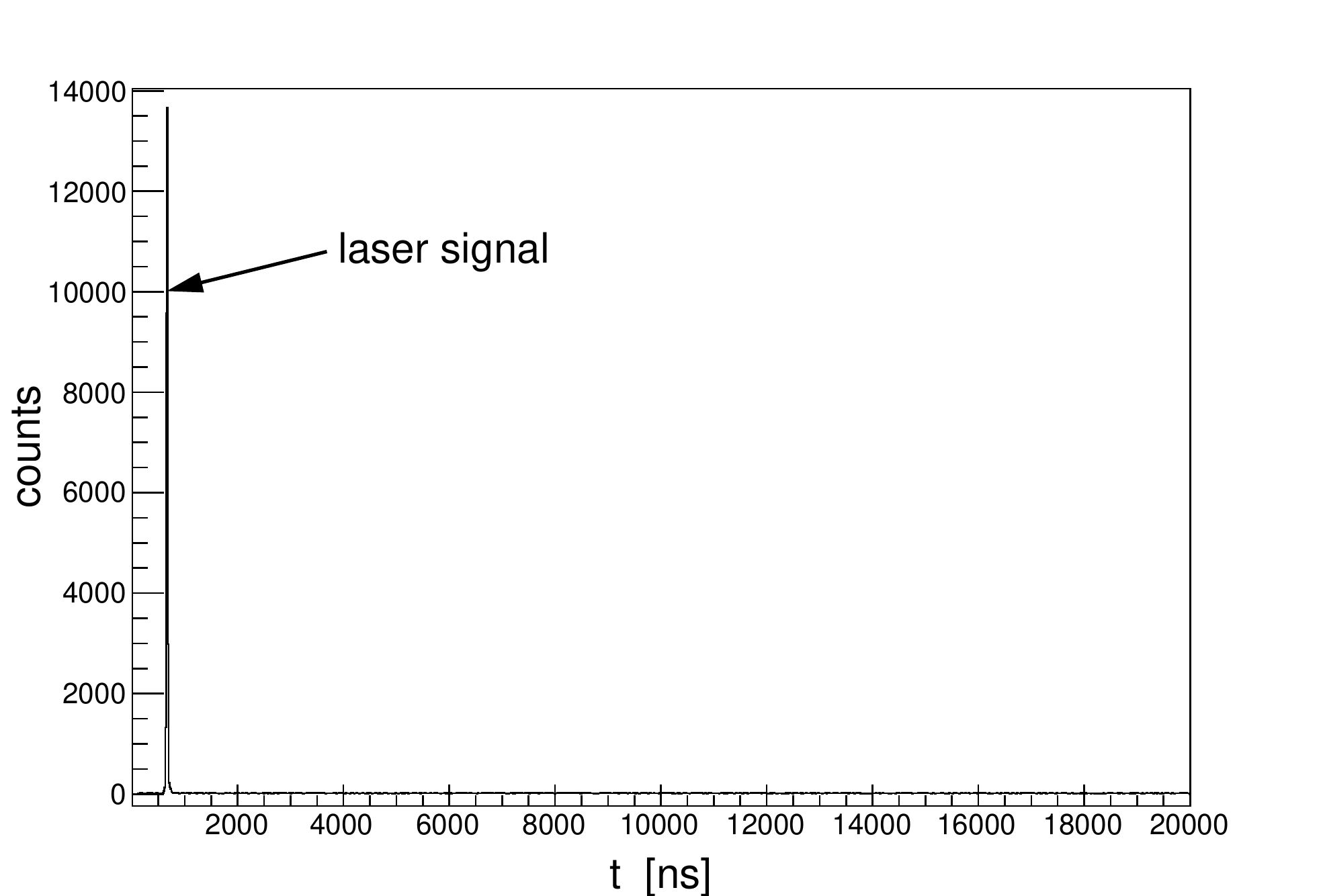}
  \label{fig_hist_raw}
  }
  \subfloat[]{
    \includegraphics[width=0.47\textwidth]{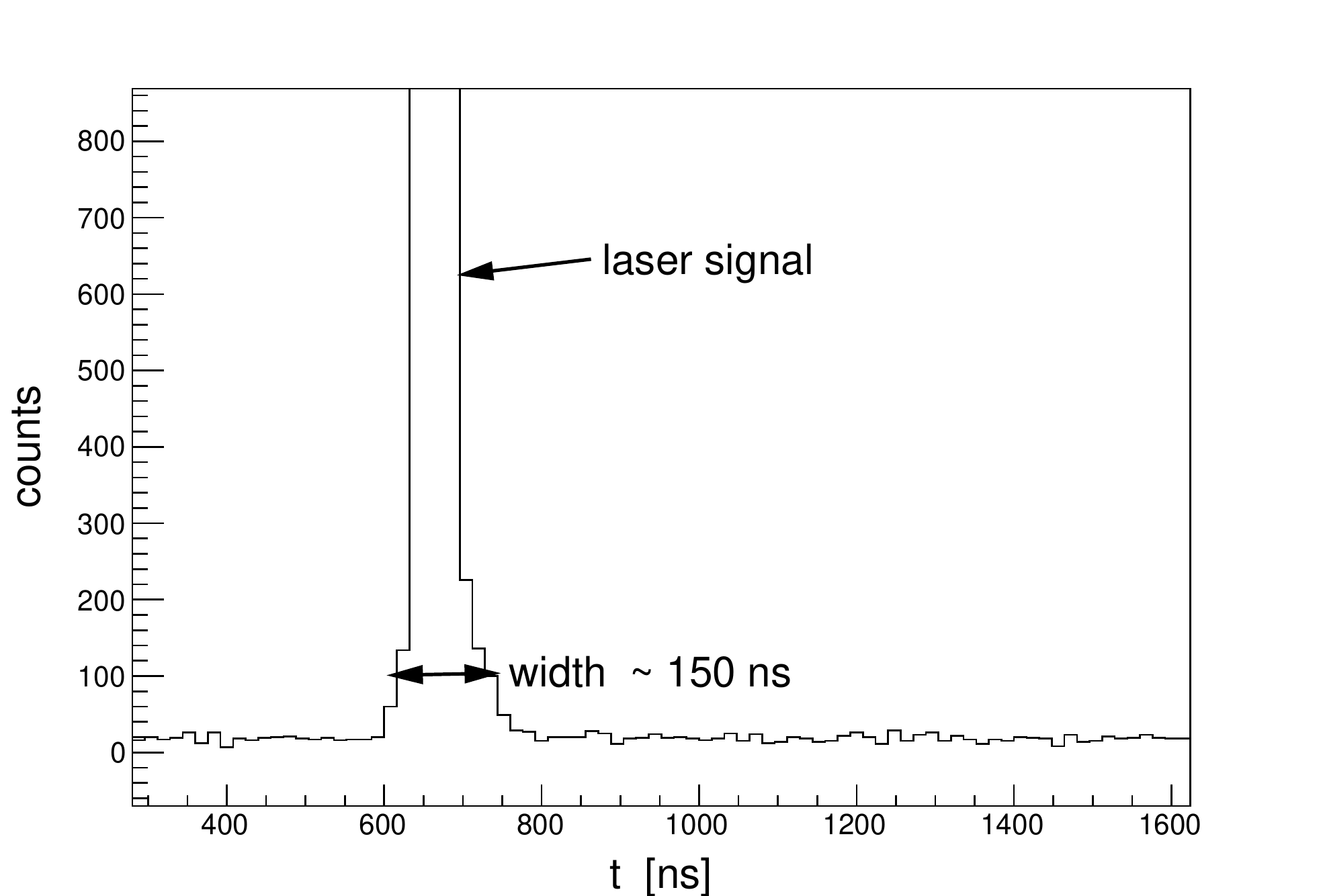}
    \label{fig_time_raw_zoom}
  }
  \caption{(a) Distribution of measured time differences between detected PMT signals and the strobe signal. (b) Zoom of figure (a). }
  \label{fig_raw_dis}
\end{figure}

The stored TB data contained all arrival time differences between all incoming signals. As a first step the time differences between the
trigger signal $t_0$ and all incoming PMT signals are plotted in figure \ref{fig_raw_dis}. Based on the TB resolution a bin width of
$\Delta t = \SI{16}{ns}$ was used for all following distributions.
\\
The raw distribution is dominated by a prominent laser signal peak. The peak maximum defines the mean arrival time of the laser signal. 
The intrinsic arrival time variation is dominated by the PMTs transit time spread of typically \SI{2.8}{ns} \cite{Flo}. 
Due to the TB resolution the measured arrival time distribution is broadened, which leads to a FWHM of the laser signal peak of \SI{16}{}-\SI{32}{ns}.
Together with a contamination of pre- and late-pulses, arriving in a time range between \SI{-30}{ns} and \SI{80}{ns} around the main signal, the total
width of the laser signal peak is about \SI{150}{ns}. The first signal within this time window is classified as the initial laser signal. 
If an initial signal is detected, all subsequent signals up to \SI{16}{\upmu s} after this initial signal are interpret as possible afterpulses 
and stored in a histogram $S_i$. 
All signals without an initial signal are treated as a distribution containing only dark noise signals and yields in a binned form the background
histogram $B_i$.\\
With a signal-to-trigger ratio $R_{\mathrm{T}}$ of about \SI{10}{\%}, only one out of 10 measurement cycles contains an initial signal plus afterpulses,
whereas the remaining \SI{90}{\%} contain only dark noise signals. 
Hence, the dark noise contribution has to be rescaled by $R_{\mathrm{T}}$ to be suitable for a dark noise correction in the initial signal distribution. 
The number of cycles with and without an initial signal determine $R_{\mathrm{T}}$, which is calculated for each PMT individually.\\
The goal of this analysis is to evaluate afterpulse effects for SPE events as the afterpulse probability increases with larger number of initial
photoelectrons (as shown in equation \ref{eq:Eap(n)}). However, the data sample is slightly contaminated by multi-photoelectron events (NPE events). 
In order to correct this effect by means of equation \ref{eq:CorrPap(n)} and \ref{eq:Eap(n)}, the average number of created photoelectrons $\overline{n}$
per initial signal has to be computed:\\ 
The probability $p_{\mathrm{pe}}(n \ge 1)$ of creating at least one photoelectron for a released trigger signal is Poisson distributed, characterized by the average
number of created photoelectrons per trigger signal $\lambda$:
\begin{equation}
p_{\mathrm{pe}}(n \ge 1) = 1 - {\rm e}^{-\lambda} = R_{\mathrm{T}} \, .
\end{equation}
This leads to an average number of created photoelectrons for each event with an initial laser signal of
\begin{equation}
\overline{n}=\frac{\lambda}{R_{\mathrm{T}}} = \frac{-{\rm ln}(1-R_{\mathrm{T}})}{R_{\mathrm{T}}} \, .
\end{equation}

%% file: sections/section_3_2_1.tex
\subsubsection{Determination of the expected number of afterpulses}\label{subsec:1am}
\begin{figure}
  \centering
  \subfloat[]{
    \includegraphics[width=0.47\textwidth]{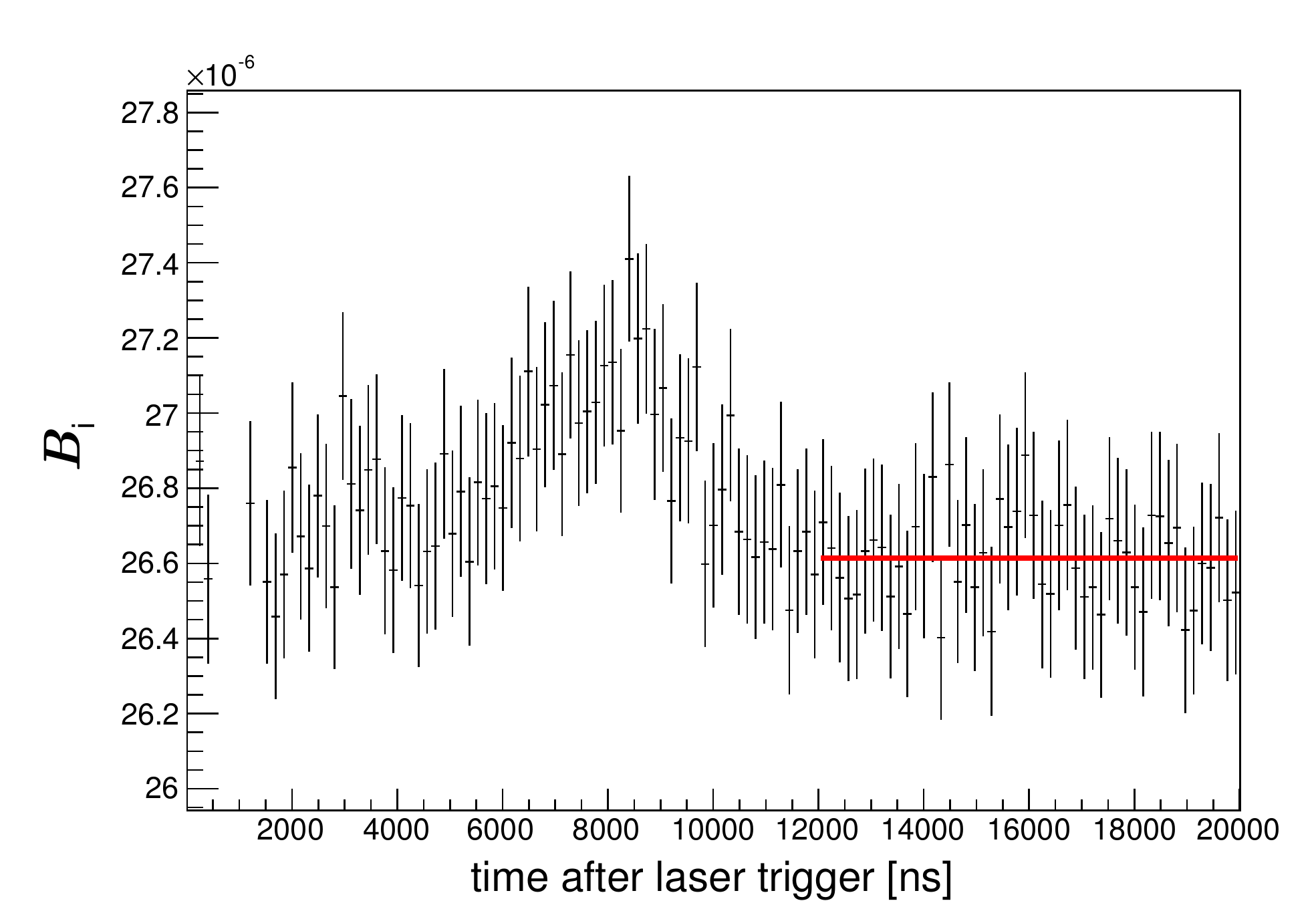}
    \label{fig_noise_dist_m1}
}
  \subfloat[]{
    \includegraphics[width=0.47\textwidth]{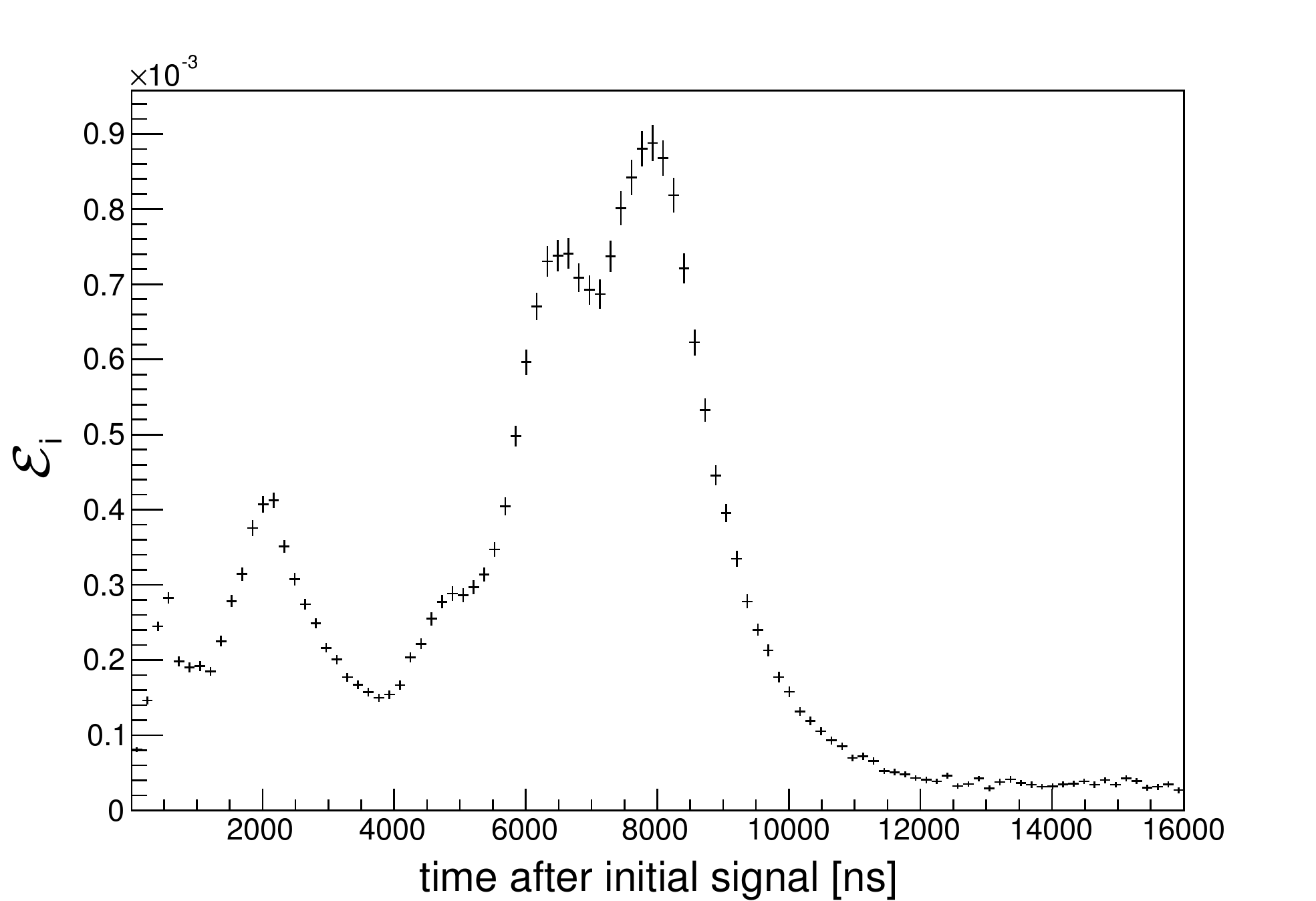}
  \label{fig_time_dist_m1}
}
  \caption{(a) Average temporal distribution of expected dark noise signals $B_i$ for all PMTs.
The vertical line indicates the flat dark noise contribution $\overline{B}$.
(b) Average temporal distribution of expected afterpulses $\mathcal{E}_i$ for all PMTs.}
  \label{fig_method1}
\end{figure}
\noindent For this analysis method all subsequent signals in the time range from \SI{0.1}{\upmu s} to \SI{16}{\upmu s} after an initial signal
are taken into account. After a PMT pulse the baseline might fluctuate causing a fake signal.
To exclude this effect as well as late-pulses, which appear up to \SI{80}{ns} following the initial pulse \cite{Flo}, a dead time of \SI{0.1}{\upmu s}
after each initial signal is applied. The dark noise distribution is assumed to be flat.
However, the observed histogram of supposed dark noise signals $B_i$ in figure \ref{fig_noise_dist_m1} shows a nonconstant contribution.
This is caused by the inefficiency of the setup: signals with a small amplitude, which do not pass any discriminator, also produce afterpulses.
These afterpulses are correlated in time with the strobe signal and contaminate the measured dark noise distribution.
In order to bypass this effect the average number of dark noise signals $\overline{B}$ was calculated in the flat region
from \SI{12}{\upmu s} to \SI{20}{\upmu s} and subtracted from the binned signal histogram $S_i$.
For the expected number of afterpulses per SPE event we yield
\begin{equation}
\label{eq:ap_m1}
\mathcal{E}_{i} = \frac{S_{i}-\overline{B}}{N_{0} \cdot \overline{n}} \, ,
\end{equation}
with the total number of initial signals $N_{0}$ and photoelectrons per initial signal $\overline{n}$. The resulting average temporal distribution
of expected afterpulses of all PMTs is shown in figure \ref{fig_time_dist_m1}.\\
\newline
Using $\mathcal{E}_{i}$, the expected number afterpulses in the time range up to \SI{16}{\upmu s} after an initial SPE signal is given by
\begin{equation}
\label{eq:aptot_m1}
\mathcal{E}^{\Delta} = \sum^{\SI{16}{\upmu s}}_{i=0} \mathcal{E}_i \, .
\end{equation}
The average number of expected afterpulses for all PMTs is
\begin{equation}
\overline{\mathcal{E}}^{\Delta} = \SI{2.68 \cdot 10^{-2}}{}\,,
\end{equation}
with a standard deviation of
\begin{equation}
\sigma = \SI{1.46 \cdot 10^{-2}}{} \, .
\end{equation}
To determine the statistical uncertainty the bin entries of $S_i$ and $B_i$ are taken to be Poisson distributed with an uncertainty
of $\sqrt{N}$. The statistical uncertainty of $\overline{n}$ is assumed to be negligible. Propagating the uncertainties using
equation \ref{eq:ap_m1} and \ref{eq:aptot_m1} yields the statistical uncertainty of $\mathcal{E}^{\Delta}$. As the statistical uncertainty
is dominated by the number of measured afterpulses, the relative statistical uncertainty
($\sigma_{\mathrm{stat}}/\mathcal{E}^{\Delta}$) is calculated for each individual PMT to allow for a better comparison among the different PMTs. 
The relative statistical uncertainty averaged over the sum of 473 PMTs is 
\begin{equation}\label{eq:errE}
\overline{\sigma}^{\mathrm{rel}}_{\mathrm{stat}} = \overline{\left( \frac{\sigma_{\mathrm{stat}}}{\mathcal{E}^{\Delta}} \right)} = \SI{5.1}{\%} \, .
\end{equation}
The only systematic effect is caused by the efficiency $\eps$ of the setup. It is defined as the ratio of the number of measured signals
$N_{\mathrm{m}}$ and the true number of signals $N_{\mathrm{t}}$. Assuming the same afterpulse probability for detected and missed signals,
the detection efficiency is given by:
\begin{equation}
\eps = \frac{N_{\mathrm{m}}}{N_{\mathrm{t}}} = \frac{N_{\mathrm{m,ap}}}{N_{\mathrm{t,ap}}} \, .
\end{equation}
$N_{\mathrm{m,ap}}$ and $N_{\mathrm{t,ap}}$ are the number of afterpulses created by the detected signals and the proper number of signals, respectively.
The efficiency can be calculated by using the the maximum ${\rm max}(\mathcal{E}_i)$ of figure \ref{fig_time_dist_m1} as an estimate
for $N_{\mathrm{m,ap}}$, whereas ${\rm max}(B_i-\overline{B})$ of figure \ref{fig_noise_dist_m1} serves as an estimate for the number of 
missed events. We thus yield
\begin{equation}\label{eq:relerrE}
\eps = \frac{ {\rm max}(\mathcal{E}_i) }{ {\rm max}(\mathcal{E}_i) + {\rm max}(B_i-\overline{B}) } = \frac{5.5}{5.505} = 0.9991\,.
\end{equation}
Compared to the statistical uncertainty this effect can be neglected.

%% file: sections/section_3_2_2.tex
\subsubsection{Determination of the afterpulse probability}\label{subsec:2am}
\begin{figure}
  \centering
  \subfloat[]{
    \includegraphics[width=0.47\textwidth]{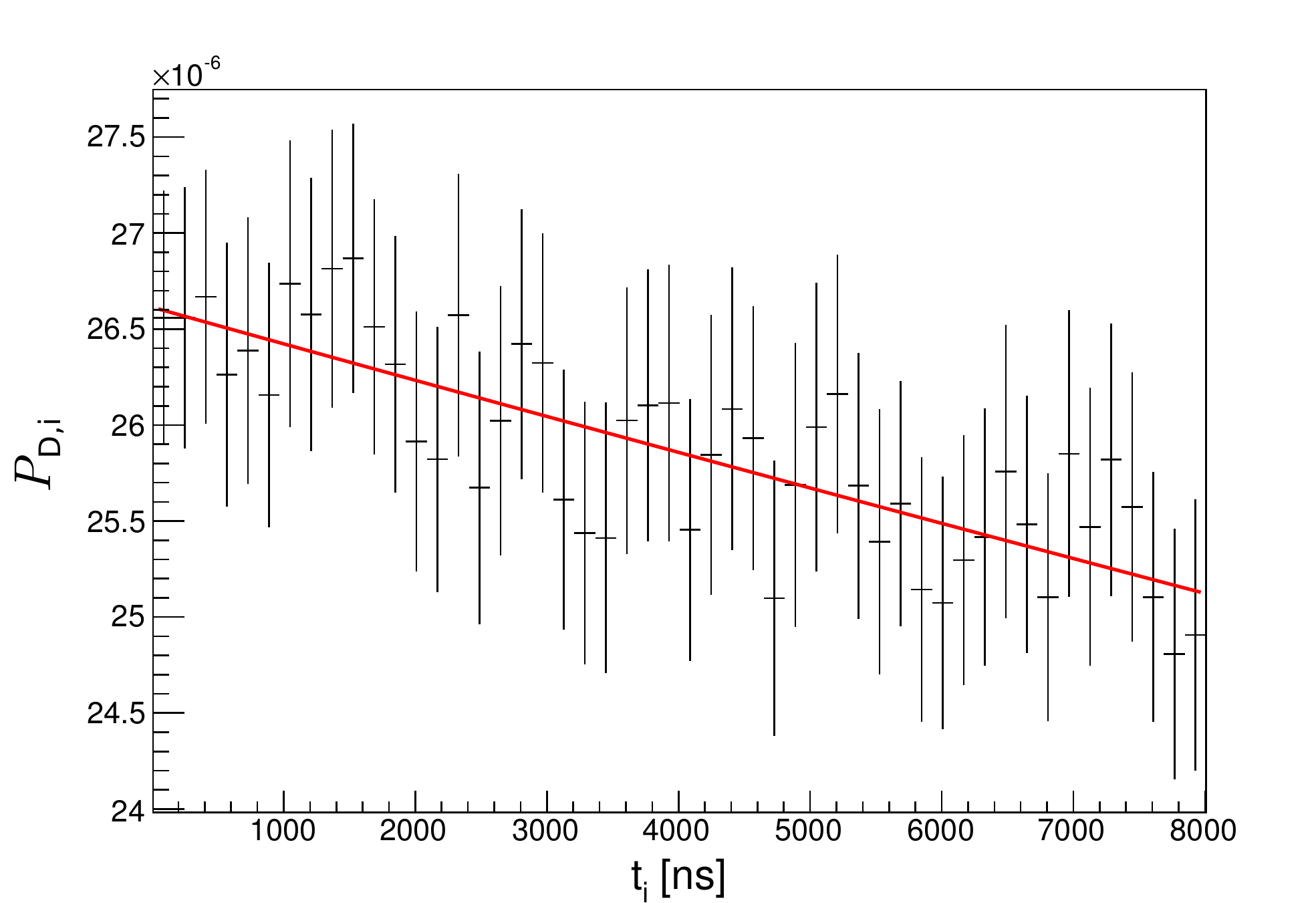}
    \label{fig_noise_dist_m2}
}
  \subfloat[]{
    \includegraphics[width=0.47\textwidth]{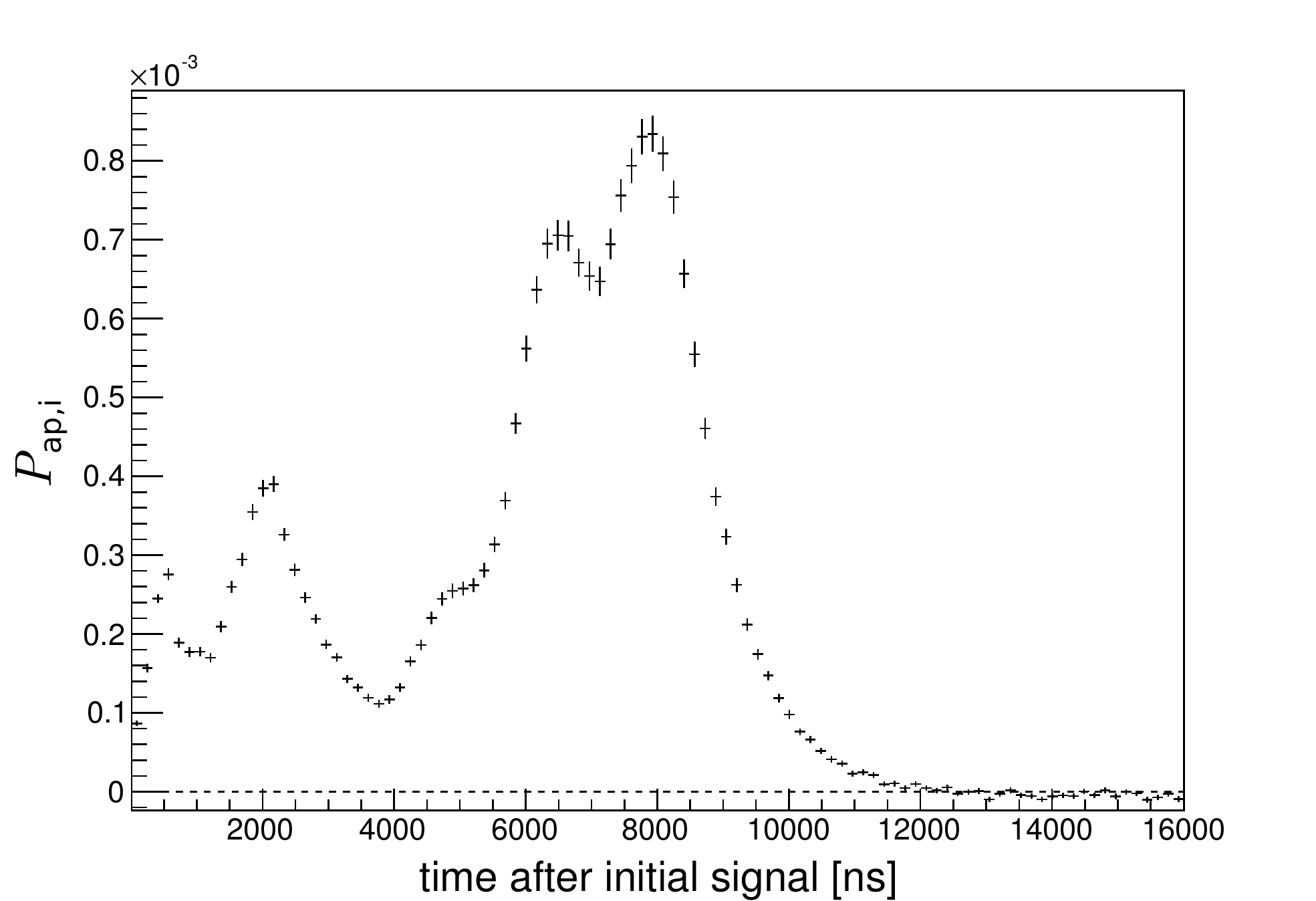}
  \label{fig_time_dist_m2}
} 
  \caption[]{(a) Average probability distribution $P_{\mathrm{D},i}= (\prod_{k=1}^{i-1}\overline{P_{\mathrm{d}}}{}_{\!,k}) \cdot P_{\mathrm{d},i}$ of measuring the first subsequent
                 dark noise signal in bin $i$ for all PMTs starting at $t_{0}=\SI{12}{\upmu s}$ after the laser trigger. The line is the assumed dark
                 noise contribution defined in equation \ref{eq:PDarkNoise}. 
(b) Average temporal afterpulse probability distribution $P_{\mathrm{ap},i}$ of all PMTs per bin width $\Delta t = \SI{160}{ns}$.}
\end{figure}
%%%%%%%%%%%%%%%%%%%%%%%%%%%%%%%%%%%%%%%%%%%%%%%5
\noindent For the second analysis the distribution $S_i$ is created containing only the first subsequent signal. 
Dividing the number of signals $S_i$ by the total number of initial signals $N_{0}$ yields the probability $P_{\mathrm{S},i}$ of measuring the first
subsequent signal in time bin $i$. $P_{\mathrm{S},i}$ is the product of the probability that no signal occurred in the time period
before bin $i$, and the probability $P_i$ of measuring at least one signal in bin $i$:
\begin{equation}
\label{eq:Pf}
P_{\mathrm{S},i} = \frac{S_i}{N_{0}} = \prod_{k=1}^{i-1}(1-P_k) \cdot P_i  \, .
\end{equation}
% The definition in equation \ref{eq:Pf} using bin-wise probabilities $P_i$ holds provided that the events are uncorrelated. This
% is given for dark noise events and first order afterpulses.
Distinguishing between afterpulses and dark noise signals the probability $P_{\mathrm{S},i}$ becomes
\begin{eqnarray}\label{eq_prob_m2}
P_{\mathrm{S},i} & = & (\prod_{k=1}^{i-1}\overline{P_{\mathrm{d}}}{}_{\!,k}) \cdot (\prod_{k=1}^{i-1}\overline{P_{\mathrm{ap}}}{}_{\!,k}) \cdot [ (1-P_{\mathrm{d},i}) \, P_{\mathrm{ap},i} + P_{\mathrm{d},i} \, (1-P_{\mathrm{ap},i}) + P_{\mathrm{d},i} \, P_{\mathrm{ap},i} ]  \nonumber \\    
{}      & = & (\prod_{k=1}^{i-1}\overline{P_{\mathrm{d}}}{}_{\!,k}) \cdot (\prod_{k=1}^{i-1}\overline{P_{\mathrm{ap}}}{}_{\!,k}) \cdot [ (1-P_{\mathrm{d},i}) \, P_{\mathrm{ap},i} + P_{\mathrm{d},i} ] \, .
\end{eqnarray}
Here, $P_{\mathrm{ap},i}$ and $P_{\mathrm{d},i}$ are the probabilities of measuring at least one first order afterpulse signal and at least one dark noise signal in time bin $i$.
$\overline{P_{\mathrm{ap}}}{}_{\!,k}$ and $\overline{P_{\mathrm{d}}}{}_{\!,k}$ are the probabilities of measuring no afterpulse signal and no dark noise signal in
each time bin $k$.\\
The dark noise probability is assumed to be Poisson distributed with a mean noise rate $R$: 
\begin{equation}
\label{eq:PDarkNoise}
\prod_{k=1}^{i-1}\overline{P_{\mathrm{d}}}{}_{\!,k} = e^{- R \cdot (i-1) \cdot \Delta t} \, ,
\end{equation}
\begin{equation}
P_{\mathrm{d},i} = 1 - e^{- R \cdot \Delta t} \, .
\end{equation}
As described in section \ref{subsec:1am} the measured dark noise distribution also contains afterpulses due to an efficiency of $\eps<1$. 
For the determination of the dark noise rate $R$ the time range between \SI{12}{\upmu s} and \SI{20}{\upmu s} after the laser trigger is considered.
Counting the number of dark noise cycles $D$ containing at least one signal between \SI{12}{\upmu s} and \SI{20}{\upmu s} and dividing
it by the total number of dark noise cycles $D_0$ leads to the probability of measuring at least one dark noise signal within \SI{8}{\upmu s}:
\begin{equation}
\frac{D}{D_0} = 1 - e^{- R \cdot \SI{8}{\upmu s}}  \, .
\end{equation}
Using this equation the rate $R$ is determined for each PMT individually.\\
The probability $P_{\mathrm{ap},i}$ is calculated in an iterative procedure:
Starting with the first bin, where the probability of measuring no previous signal is equal to one, the probability $P_{\mathrm{ap},1}$ is given by
\begin{equation}
P_{\mathrm{ap},1} = ( P_{\mathrm{S},1} + e^{-R \cdot \SI{16}{ns} } - 1 ) \cdot e^{ R \cdot \SI{16}{ns} }   \, .
\end{equation}
Using this result the probability of the following bin $P_{\mathrm{ap},2}$ can be calculated.
Repeating this procedure up to $t=\SI{16}{\upmu s}$ the probability $P_{\mathrm{ap},i}$ is calculated for each time bin.\\
To yield an afterpulse probability per SPE event, the bin-wise probabilities are corrected by the average number of initial photoelectrons
$\overline{n}$ using equation \ref{eq:CorrPap(n)}:
\begin{equation}
\mathcal{P}_{i} = 1 - \big( 1-P_{\mathrm{ap},i}(\overline{n}) \big)^{1/\overline{n}} \, .
\end{equation}
The average temporal afterpulse probability distribution for all PMTs is shown in figure \ref{fig_time_dist_m2}.\\ 
Since the quantities $P_{\mathrm{ap},i}$ are bin-wise probabilities, the sum over several bins does not represent any physical quantity anymore.
However, the afterpulse probability $\mathcal{P}$ can be calculated by taking the advantage of 
\begin{equation}
\prod^{\SI{16}{\upmu s}}_{i = 0} (1-\mathcal{P}_{i})
\end{equation}
being the probability of measuring no afterpulse event in the time range from 0 to \SI{16}{\upmu s}.
For the afterpulse probability of measuring at least one afterpulse in this time interval we can conclude that
\begin{equation}
\label{eq:aptot_m2}
\mathcal{P} = 1 - \prod^{\SI{16}{\upmu s}}_{i = 0} (1-\mathcal{P}_{i}) \, .
\end{equation}
Figure \ref{fig_hist_AP_prob_m2} shows the total afterpulse probability $\mathcal{P}$ of all PMTs in a histogram. 
For the average afterpulse probability we yield
\begin{equation} 
\overline{\mathcal{P}} = \SI{2.27 \cdot 10^{-2}}{}
\end{equation}
with a standard deviation of
\begin{equation} 
\sigma = \SI{1.22 \cdot 10^{-2}}{} \, .
\end{equation} 
Since $\mathcal{P}$ is determined in an iterative procedure, the statistical uncertainty is calculated for each time bin individually, while the
errors of the preceding bins are propagated to the next bin. The statistical uncertainty of the entries of the initial signal and dark noise signal
distribution are assumed to be Poisson distributed with an uncertainty of $\sqrt{N}$. The statistical uncertainty of the average number of
photoelectrons $\overline{n}$ per initial signal is again assumed to be negligible. Propagating these uncertainties via equation \ref{eq_prob_m2}
and computing the squared sum over all $\sigma_{\mathrm{stat},i}$ yields the statistical uncertainty of $\mathcal{P}$. As the statistical uncertainty
is dominated by the number of afterpulses, the relative statistical uncertainty ($\sigma_{\mathrm{stat}}/\mathcal{P}$) is calculated for each PMT,
which simplifies the comparison of different PMTs' results. The relative statistical uncertainty averaged over the sum of all PMTs is 
\begin{equation}
\overline{\sigma}^{\mathrm{rel}}_{\mathrm{stat}} = \overline{\left( \frac{\sigma_{\mathrm{stat}}}{\mathcal{P}} \right)} = \SI{5.4}{\%} \, .
\end{equation}
Although each uncertainty $\sigma_{\mathrm{stat},i}$ is propagated from bin to bin, the achieved uncertainty is only slightly lager compared to the
relative error of $\mathcal{E}^{\Delta}$ (cf. equation \ref{eq:errE}).\\ 
The only systematic effect is caused by the efficiency of the setup, as $\eps<1$. 
Due to high complexity of the calculation, impact of this effect on our result was not computed.
Nevertheless, we can assume that the effect's size is of the same order as in equation \ref{eq:relerrE}, and therefore neglect the systematic uncertainty.

%% file: sections/section_3_2_3.tex
\subsubsection{Connection between both analyses' results}\label{subsec:comp}
\begin{figure}
  \centering
  \subfloat[]{
    \includegraphics[width=0.47\textwidth]{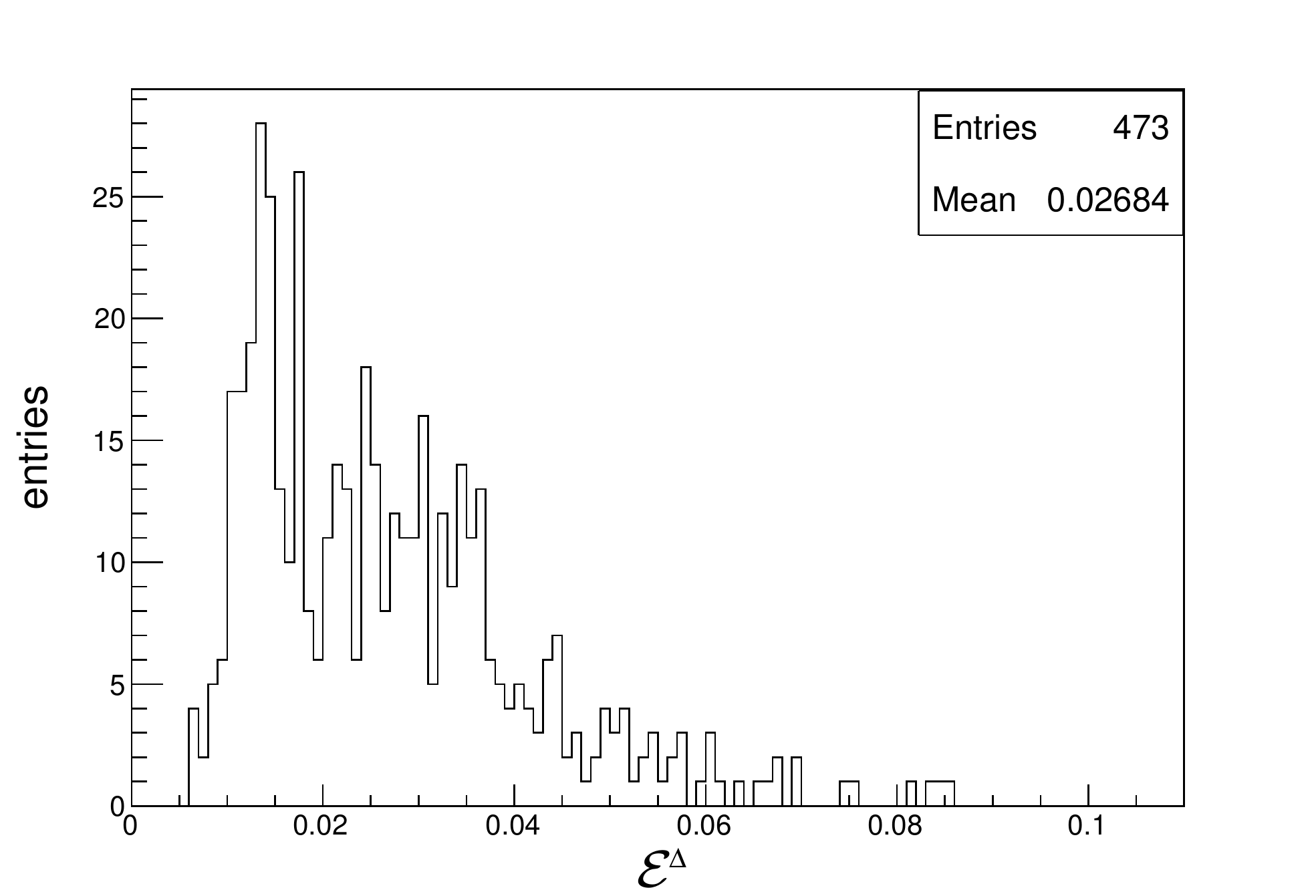}
    \label{fig_hist_AP_prob_m1}
}
  \subfloat[]{
    \includegraphics[width=0.47\textwidth]{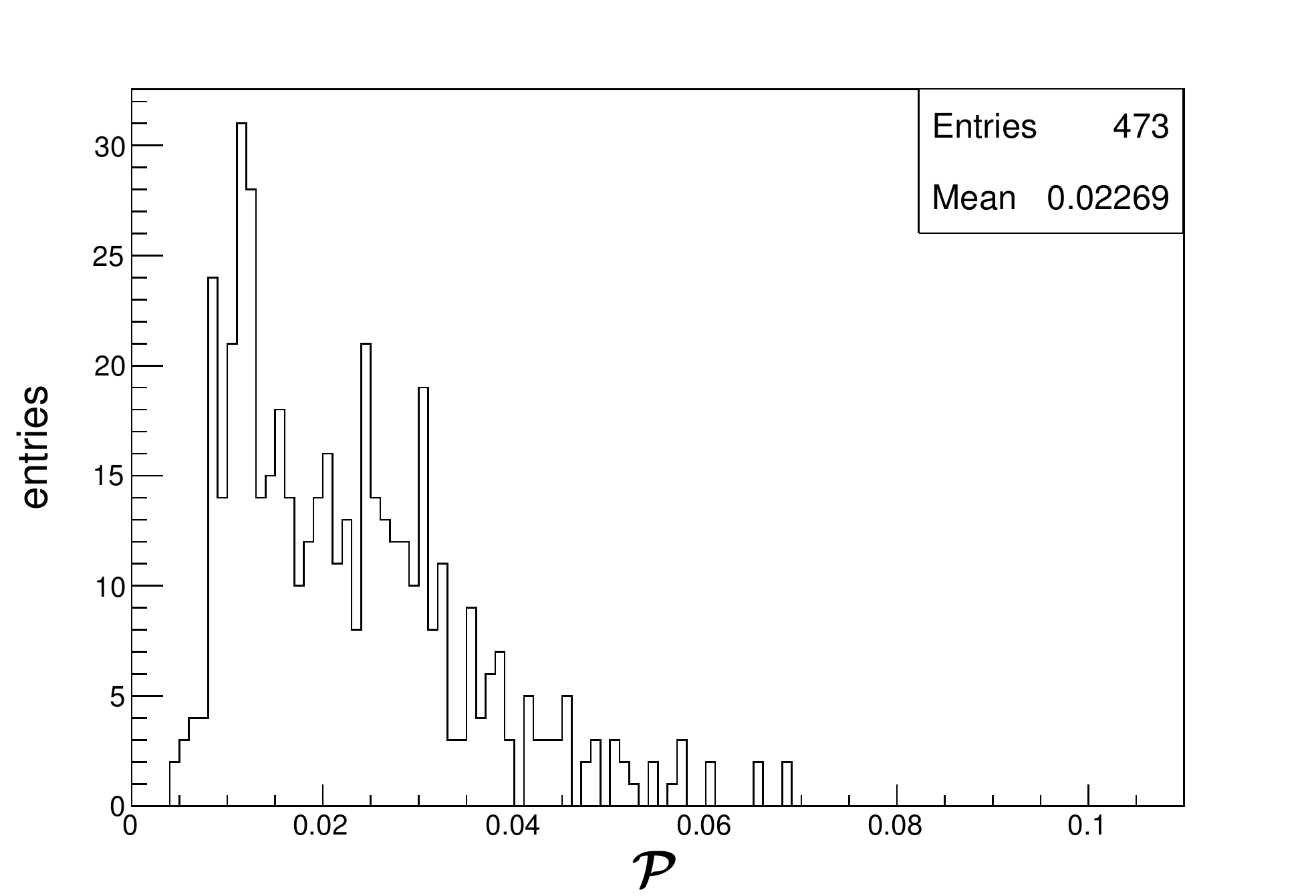}
  \label{fig_hist_AP_prob_m2}
}
  \caption[]{(a) Distribution of the expected number of afterpulses $\mathcal{E}^{\Delta}$ (equation \ref{eq:aptot_m1}) for all PMTs. 
 (b) Distribution of the afterpulse probability $\mathcal{P}$ (equation \ref{eq:aptot_m2}) for all PMTs.}
  \label{fig_hist_AP_prob}
\end{figure}

The values of $\mathcal{E}^{\Delta}$ as well as $\mathcal{P}$ are smaller than 0.1 for all PMTs, as shown in figures \ref{fig_hist_AP_prob_m1}
and \ref{fig_hist_AP_prob_m2}. This means not only the expected number of afterpulses per SPE is less than 0.1, but also probability for a single PE to
obtain at least one afterpulse is smaller than 10\,\%. Therefore the selection specification of the Double Chooz experiment were well fulfilled, which
requested an afterpulse probability of less than 10\,\% in the time range of 0 to \SI{16}{\upmu s}.
\\
\begin{figure}
  \centering
  \subfloat[]{
  \includegraphics[width=0.47\textwidth]{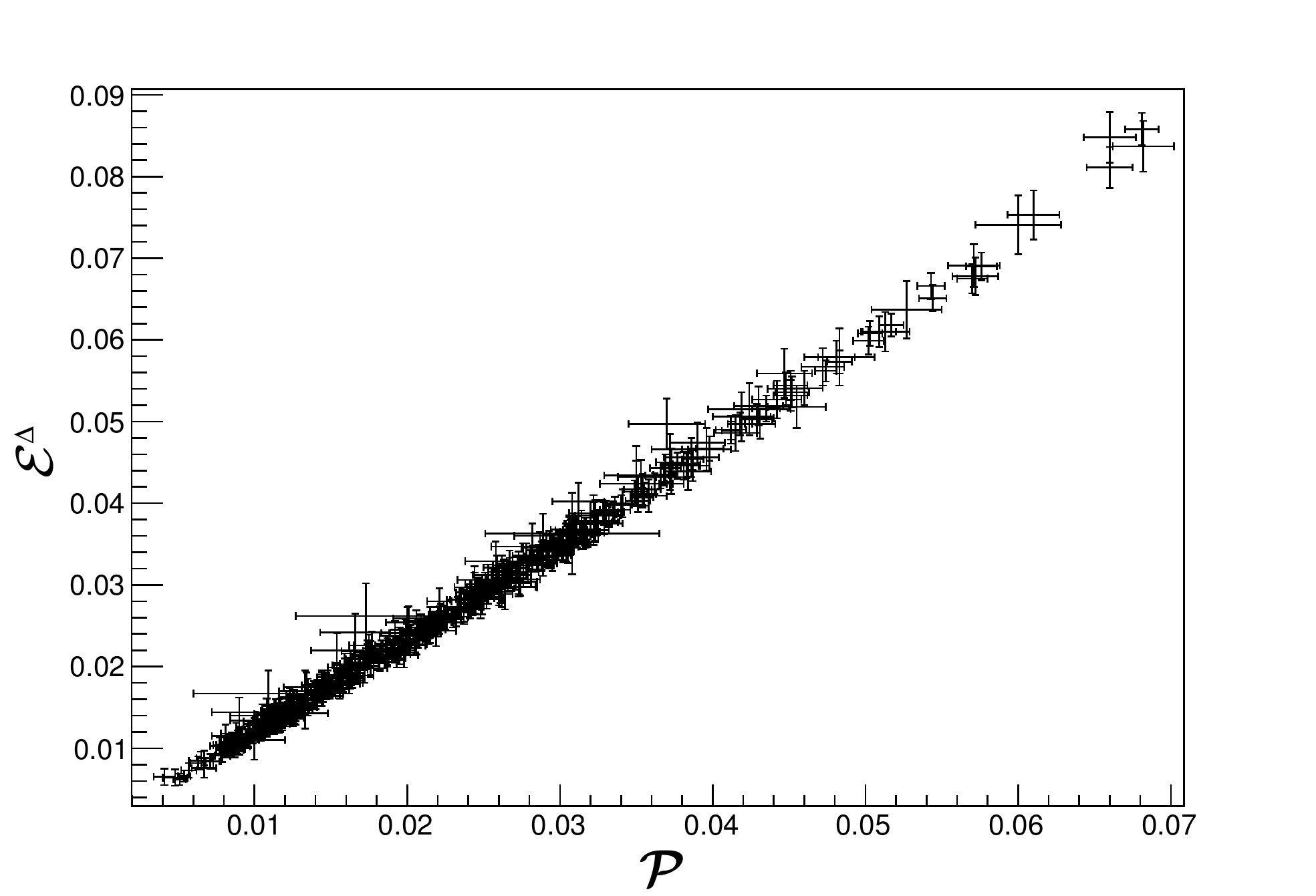}
  \label{fig_total_ap_bm}
}
  \subfloat[]{
  \includegraphics[width=0.47\textwidth]{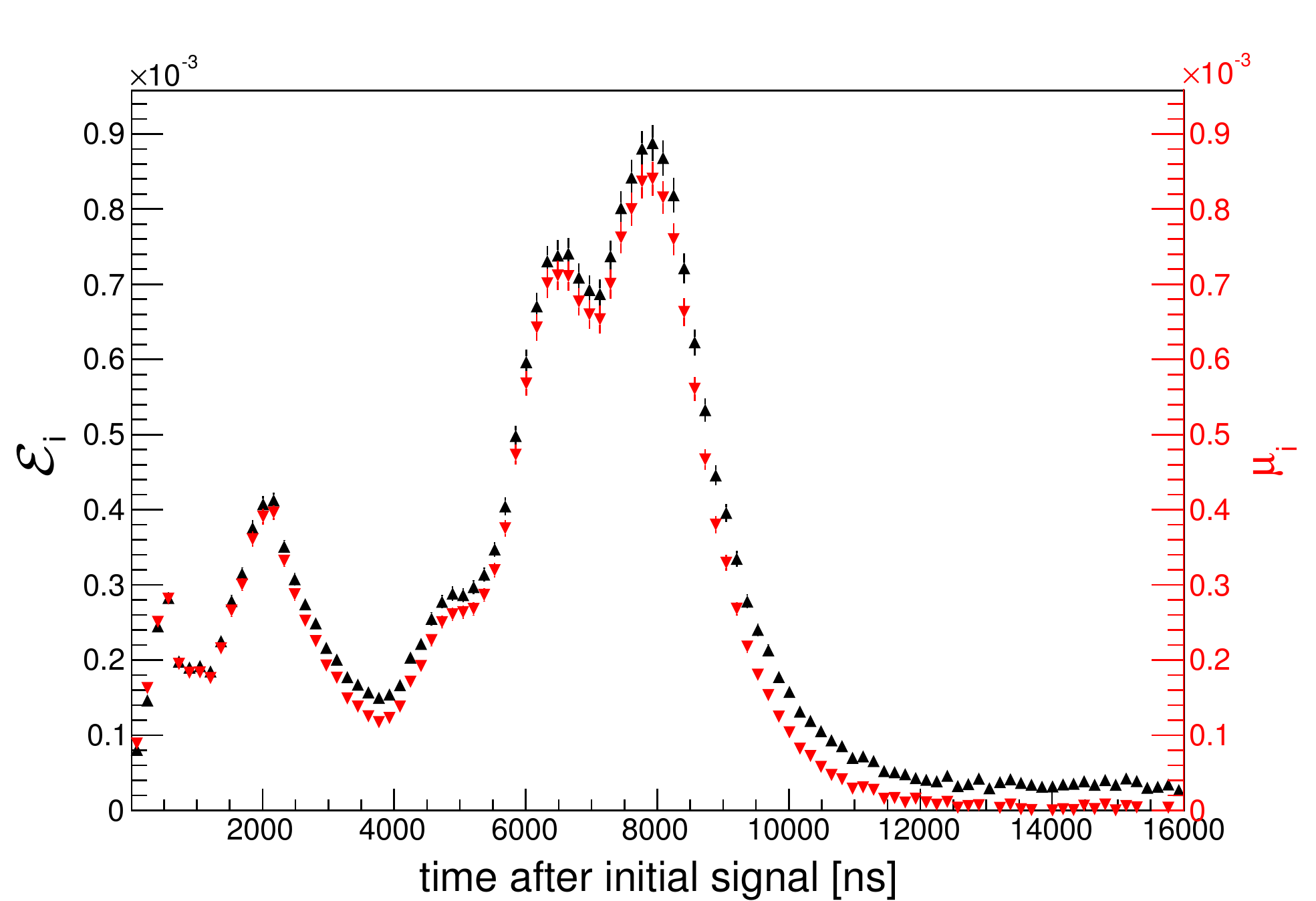}
  \label{fig_time_dist_bm}
}
  \caption{(a) $\mathcal{E}^{\Delta}$ vs.~$\mathcal{P}$ for all PMTs.
   (b) Temporal distributions of the expected number of afterpulses $\mathcal{E}^{\Delta}$ and the afterpulse probability $\mu_{i}$, averaged
       over the sum of 473 PMTs.}   
  \label{fig_comparison} 
\end{figure}
\\
The correlation between $\mathcal{E}^{\Delta}$ and $\mathcal{P}$ is plotted in figure \ref{fig_total_ap_bm}.
In figure \ref{fig_time_dist_bm} we can see that the temporal distribution $\mu_i$ (gained via $\mathcal{P}_i$ together with
equation \ref{eq:CorrPap(n)} and equation \ref{eq:APProb}) approaches zero for $t>\SI{12}{\upmu s}$, i.e.~no further
first order afterpulses occur after \SI{12}{\upmu s}. Thus, the calculated probability $\mathcal{P}$ includes all created first order afterpulses.
On the other hand, the temporal distribution $\mathcal{E}_i$ does not approach zero for times larger than \SI{16}{\upmu s}. Therefore
the expected number of afterpulses $\mathcal{E}^{\Delta}$ is calculated by summing $\mathcal{E}_i$ from \SI{0}{} to \SI{16}{\upmu s},
including only a fraction $f$ of higher order afterpulses.
This observation shows, that the fraction $f$ has to be considered in our calculations.\\
According to equations \ref{eq:APProb} and \ref{eq:Eap0-16}, the average number of photoelectrons $\overline{\ell}$ contained by an afterpulse can be calculated
using $\mathcal{E}^{\Delta}$ and $\mathcal{P}$:
\begin{equation}
\overline{\ell} = - \frac{ \mathcal{E}^{\Delta} + {\rm ln}(1-\mathcal{P})}{\mathcal{E}^{\Delta} \cdot {\rm ln}(1-\mathcal{P}) + ({\rm ln}(1-\mathcal{P}))^2 \cdot (1-f)} \, .   
\end{equation}
The fraction $f$ is calculated using the temporal distribution $E_{\mathrm{h},i}$ of expected higher order afterpulses (cf. equation \ref{eq:fint}).
To determine $E_{\mathrm{h},i}$ via equation \ref{eq:Eho(mu,ti)} we use the average temporal distributions of $\mathcal{E}_i$ and $\mu_i$
(cf. figure \ref{fig_time_dist_bm}). This leads to an average fraction of
\begin{equation} 
\overline{f} = \SI{82.6}{\%} \, .
\end{equation}
Compared to the uncertainties of the other values used to calculate $\overline{\ell}$, the uncertainty of $\overline{f}$ can be neglected.\\
Using $\overline{f}$, $\mathcal{E}^{\Delta}$ and $\mathcal{P}$, $\overline{\ell}$ is calculated for each PMT individually.
For the error weighted mean of all PMTs we receive
\begin{equation}
\overline{\ell} = \SI{(4.95 \pm 0.13)}{PE} \, .
\end{equation}

%% file: sections/section_4.tex
\section{Timing and charge measurements for dedicated PMTs}
In order to determine the temporal occurrence of afterpulses correlated with their charge, an additional and more detailed measurement was realized. 
Due to complexity, these tests were performed only for a few PMTs. 
To gain information about the dependency of the afterpulse properties on the incident light levels, different intensities were used.

%% file: sections/section_4_1.tex
\subsection{Experimental Setup}
\begin{figure}[t]
 \centering
 \includegraphics[width=0.95\textwidth]{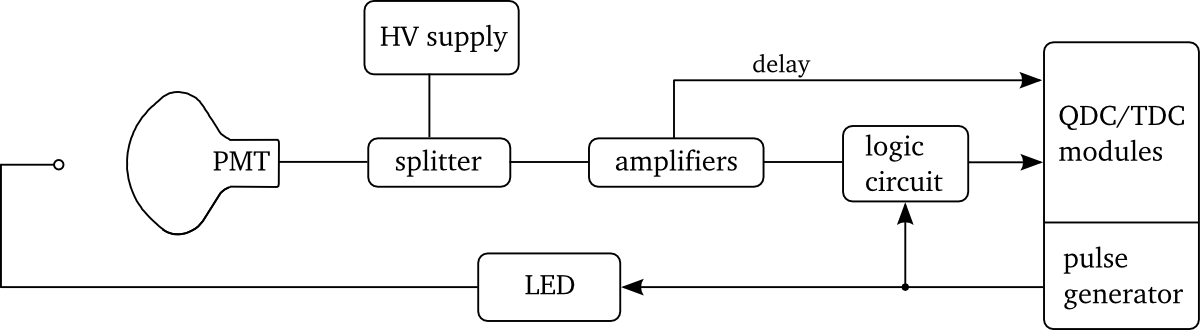}
 \caption{Scheme of the experimental setup for detailed measurements.}
 \label{f_hd1}
\end{figure}

The tested PMTs were arranged in a light-tight Faraday room as described above. All electronics components were installed outside the room,
the setup is shown in figure \ref{f_hd1}.
A custom-built LED board with a wavelength of $\SI{380}{nm}$ served as light source. 
To allow for a simultaneous measurement of timing and charge of the afterpulses, both a time-to-digital converter (TDC, model V775/ CAEN) and a
charge-to-digital converter (QDC, model V792/ CAEN) were used. 
Another VME module producing rectangular pulses with a distinct frequency, acted as pulse generator.
It provided several output channels and the pulses of each output could be adjusted individually in width and in their delay with respect to the
other signals. In combination with discriminator modules and a logical circuit composed of AND/OR and gate delay modules, pulses occurring after
an initial PMT pulse could be measured. 
%%%%
%%%%
\\
The TDC used for this measurement was a single-hit TDC with a resolution of \SI{0.3}{ns} per bin.
It records the first appearing afterpulse in a time window of approximately $\SI{1200}{ns}$. 
Therefore several measurements at delay steps of $\SI{1}{\upmu s}$ were carried out.
Due to baseline distortion effects caused by the initial PMT pulses, the search for afterpulses was started $\SI{375}{ns}$ after an initial pulse.
Assuming that the dark noise is independent in time, the measured time and charge spectra of the afterpulses are corrected by taking an additional
dark rate measurement with a time delay of $\sim \SI{70}{\upmu s}$ with respect to the initial pulse.

%% file: sections/section_4_2.tex
\subsection{Analysis and Results} \label{hd_vme}
\begin{figure}
 \centering
 \subfloat[]{\label{f_hd2a}\includegraphics[width=0.47\textwidth]{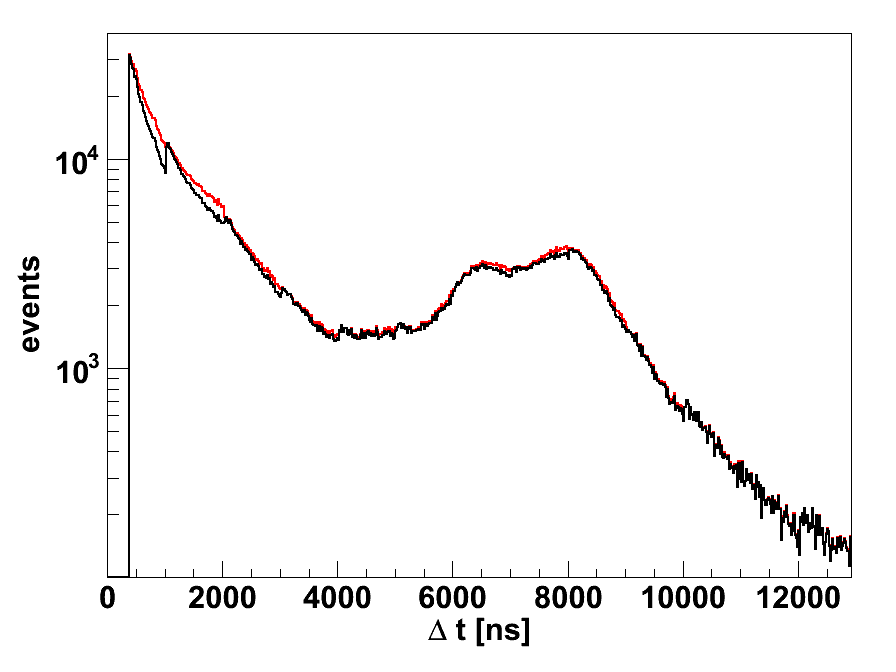}}
 \,
 \subfloat[]{\label{f_hd2b}\includegraphics[width=0.47\textwidth]{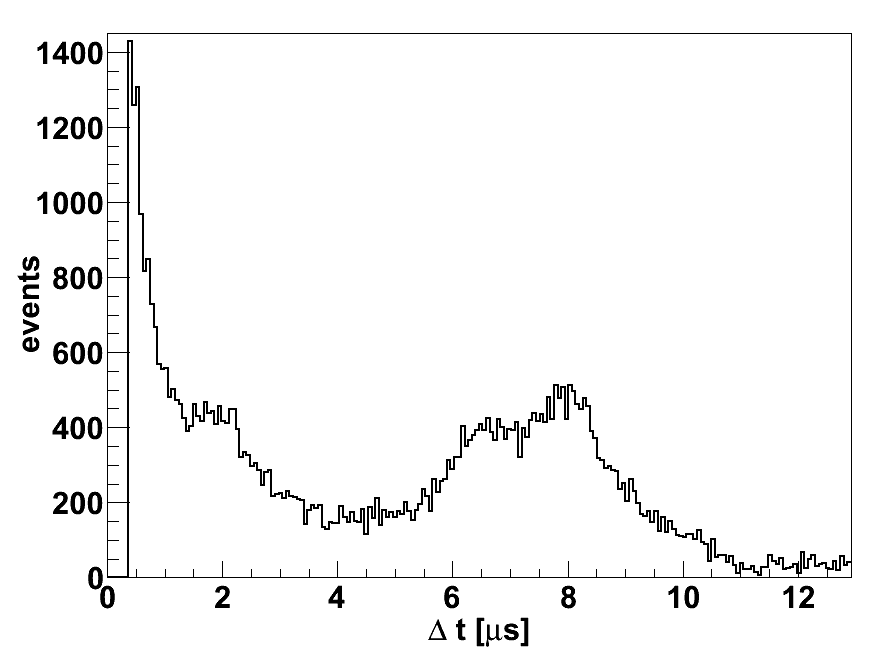}}
 \caption{(a) Afterpulse timing spectrum without dark noise correction $S_i$ (corrected spectrum $S^{\mathrm{corr}}_i$ in red) for $\overline{n} = \SI{17.4}{PE}$.
	  (b) Afterpulse timing spectrum with dark noise correction for $\overline{n} = \SI{1.1}{PE}$}
\end{figure}

\begin{figure}
 \centering
 \subfloat[]{\label{f_hd3a}\includegraphics[width=0.49\textwidth]{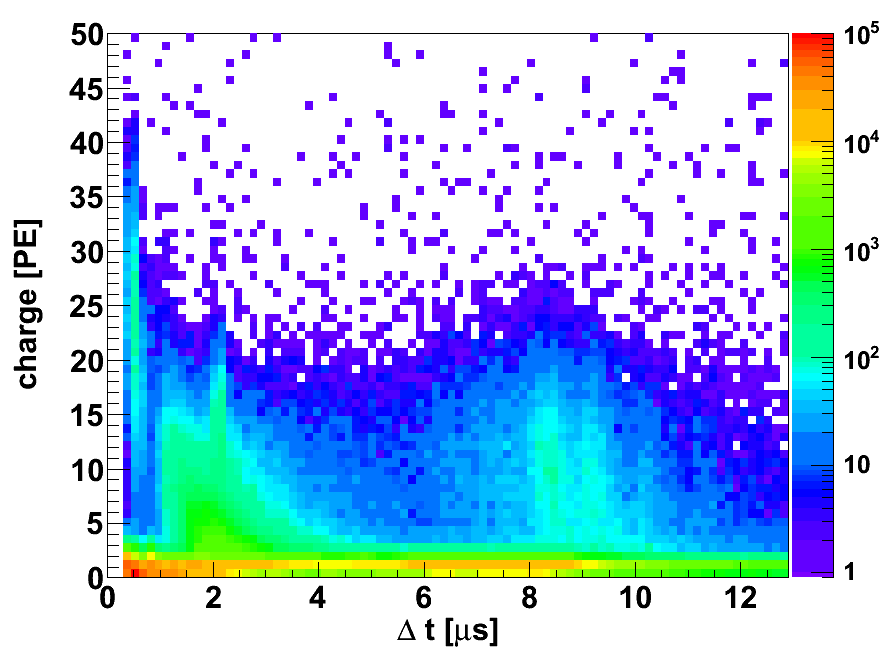}}
 \,
 \subfloat[]{\label{f_hd3b}\includegraphics[width=0.49\textwidth]{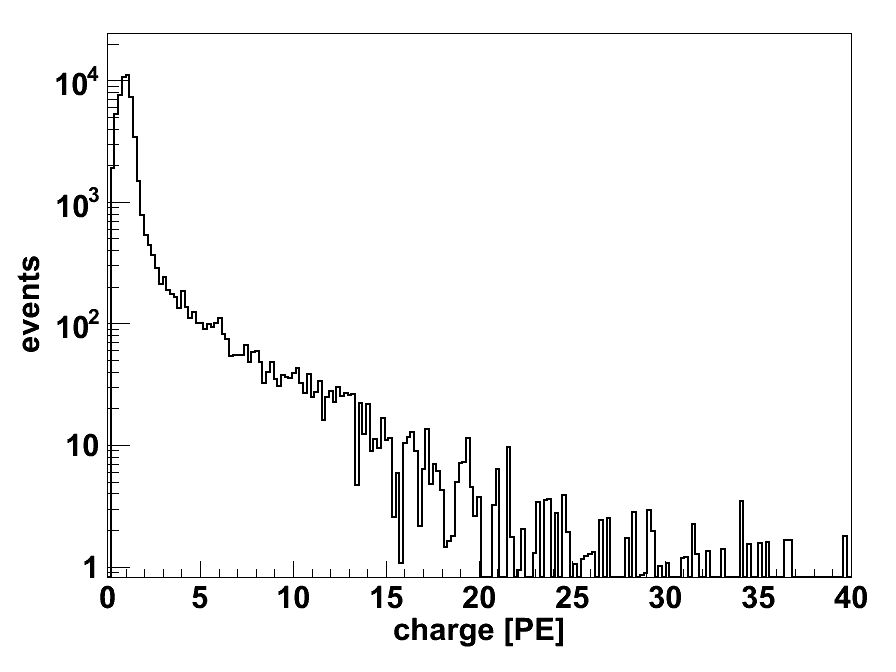}}
 \\
 \subfloat[]{\label{f_hd3c}\includegraphics[width=0.49\textwidth]{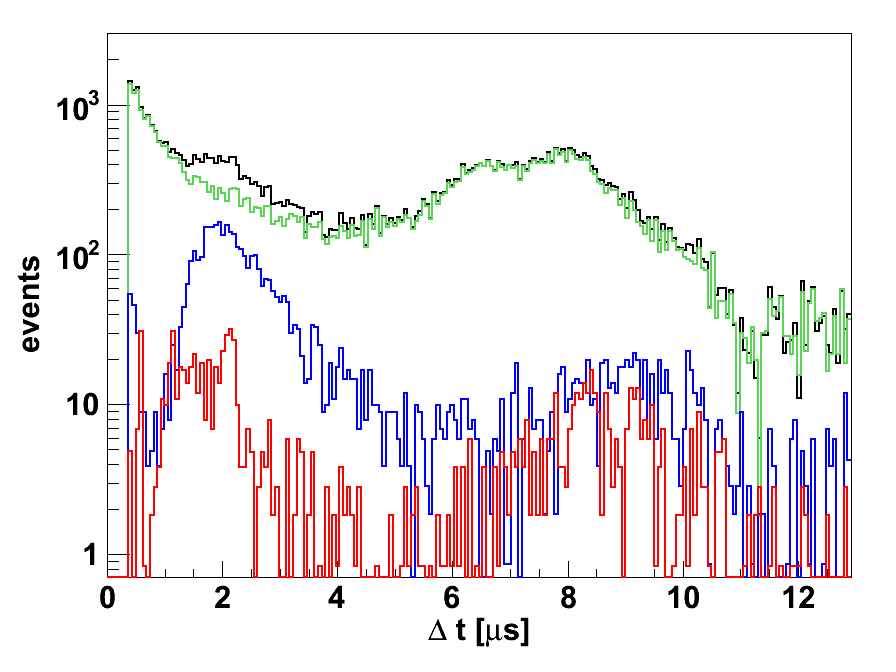}}
 \,
 \subfloat[]{\label{f_hd3d}\includegraphics[width=0.49\textwidth]{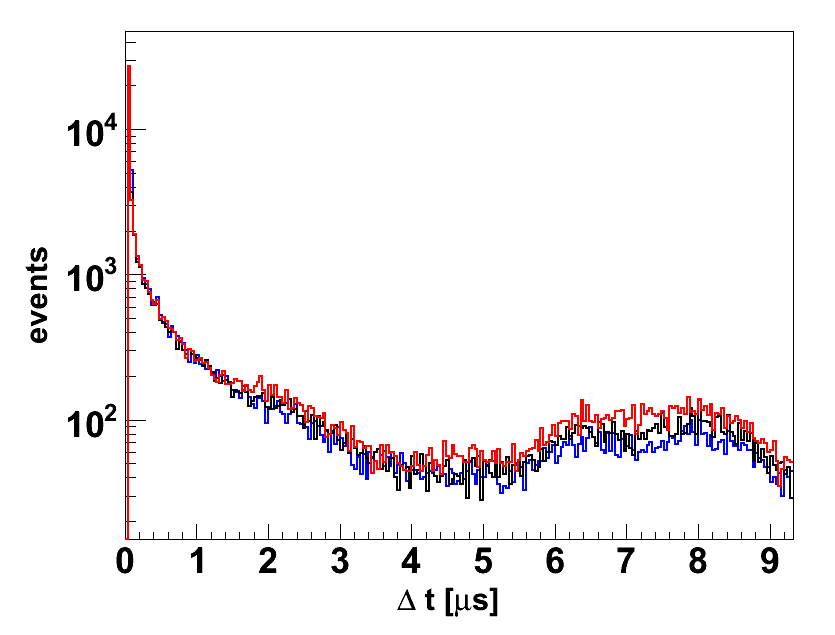}}

  \caption{(a) Scatter plot $(t,\ell)$, temporal occurrence $t$ vs. the charge $\ell$ of the afterpulses for $\overline{n} = \SI{17.4}{PE}$.
           (b) Charge spectrum of the afterpulses for $\overline{n} = \SI{1.1}{PE}$.
	   (c) Afterpulse timing spectrum with charge cuts for $\overline{n} = \SI{1.1}{PE}$; black: complete spectrum,
                 green: $0<\ell<\SI{2}{PE}$, blue:~$2<\ell<\SI{10}{PE}$, red:~$10<\ell<\SI{25}{PE}$.
	   (d) Afterpulse timing spectrum for $\overline{n} = \SI{14.8}{PE}$ measured with different high voltages:
		 blue: $\mathrm{HV}_{0.5} = \SI{1130}{V}$,
		 black: $\mathrm{HV}_1 = \SI{1229}{V}$,
		 red: $\mathrm{HV}_2 = \SI{1340}{V}$}.
\label{f_hd3}
\end{figure}

The results presented in this section characterize the afterpulse properties of one particular PMT in the range $\Delta = [\SI{375}{ns};\SI{12}{\upmu s}]$
after the main PMT pulses, as the afterpulse behaviour turned out to be similar for all tested PMTs.\\
\newline
Due to usage of a single-hit TDC several individual measurements covering different regions after a main pulse have been taken.
Merging the datasets results in steps at each edge to a subsequent measurement range.
This happens since events in earlier bins suppress possible afterpulse events in any of the subsequent bins of one independent measurement.
These kinks can be removed by the following bin-by-bin correction of the spectra.
The correction assumes that the probabilities to obtain afterpulses in certain bins do not depend on each other and the maximum expected number
of afterpulses per bin is smaller than 1. 
This is implied by the appropriate assumption that multiple afterpulse events produced by a single photoelectron are independent events and that the
contribution of higher order afterpulses within a measurement interval of $\SI{1}{\upmu s}$ is negligible.
Coincidental afterpulses in the same bin can be excluded due to the low afterpulse probability in a bin width of $\SI{0.3}{ns}$.\\
The first bin of a dataset does not need to be corrected. 
The corrected number of entries for the second bin is given by
%%%%
\begin{equation} \label{eq_hd1}
S^{\mathrm{corr}}_{2} = \frac{S_{2}}{1 - S_{1}/N_{0}}\,,
\end{equation}
%%%%
where $N_{0}$ is the number of initial pulses.
The correction for higher bins accounts for the capability of one or more events in any prior bin to suppress afterpulse events in the bin of interest.
Hence, the number of events of the i-th bin $S^{\mathrm{corr}}_{i}$ is
%%%%
\begin{equation} \label{eq_hd2}
S^{\mathrm{corr}}_{i}  = S_{i} \cdot \left(\prod_{k=1}^{i-1} (1-S^{\mathrm{corr}}_{k}/N_{0}) \right)^{-1} \,.
\end{equation}
Figure \ref{f_hd2a} shows both the measured temporal afterpulse spectrum in black and the corrected temporal spectrum in red.
The additional dark rate measurement has as well been corrected via equation \ref{eq_hd2} and was used to calculate the mean number of
dark rate signals per bin $\overline{B}$. The spectrum $S^{\mathrm{corr}}_{i}$ was dark noise corrected in each bin by subtraction of $\overline{B}$.
\\
\newline
%%%% Results %%%%
The histogram with all time differences between initial pulse and afterpulses is plotted in figure~\ref{f_hd2b}, representing the afterpulse timing spectrum
for the measurement with $\overline{n} = \SI{1.1}{PE}$. Distinctive peaks can be found at around $2$, $6.5$ and $\SI{8}{\upmu s}$.
Comparing the afterpulse timing spectra, the trigger board (figure \ref{fig_time_dist_bm}) and the TDC measurement are in agreement for times greater
than $\SI{4}{\upmu s}$, for smaller times the TDC measurement shows an additional exponential contribution. To investigate this observation further,
a control measurement using the setup shown in figure \ref{f_hd1} with a different type of PMT was performed. As expected a differing peak structure
but also an identical exponentially decreasing part was observed in the timing spectrum.
Furthermore, measurements using different LED intensities could demonstrate a non-linear dependence between the exponential component and the LED intensity.
This led to the conclusion that the exponential contribution is induced by afterglowing effects of the LED.
\\\\
The afterpulse charge spectrum (figure \ref{f_hd3b} and \ref{f_hd3b}) is dominated by small charges of a few photoelectrons.
However, a contribution of higher charged afterpulses up to $\SI{40}{PE}$ also becomes evident.
The exponential tail in the timing due to LED afterglow is associated with charges $\ell<\SI{2}{PE}$, as shown in figure~\ref{f_hd3c}.
Also, afterpulses with a timing of $5$-$\SI{11}{\upmu s}$ contain a considerable amount of low charges. 
Regions of afterpulses with higher charges do not fully correspond to the peaks seen in the time spectrum in figure \ref{f_hd2b}. 
In figure \ref{f_hd3a} and \ref{f_hd3c} regions of afterpulses transferring large charges can be identified at $\SI{500}{ns}$, from $1.5$ to
$\SI{2}{\upmu s}$ and around $\SI{8.5}{\upmu s}$.\\
The effect of suppressing possible subsequent pulses due to usage of a single-hit TDC can not be corrected
in case of the charge spectra (cf. equation \ref{eq_hd2} for the time spectra).
Hence, a slight deformation of each charge spectrum is expected.\\
Measurements with higher light levels confirmed the assumption that highly charged afterpulses arise due to ion feedback.
If more light is focused on the photocathode, and thus a larger number of initial photoelectrons $n$ is released, the number of afterpulses rise,
but the spectral shape of timing and charge do not change.
\\\\
Furthermore, additional afterpulse measurements with three different voltages were performed for the same PMT for $\overline{n} = \SI{14.8}{PE}$.
The conventional high voltage $\mathrm{HV}_1$ leads to a gain of $\SI{0.8}{pC}$ per photoelectron, $\mathrm{HV}_2$ provides a factor two in
gain ($\SI{1.6}{pC}$ per photoelectron), and $\mathrm{HV}_{0.5}$ results in half the gain of $\mathrm{HV}_1$. 
No significant shift of the peaks structure could be observed.
The change in high voltage showed only a slight effect on the number of afterpulses as seen in figure \ref{f_hd3d}.
%%%%
%%%%

%% file: sections/section_5.tex
\section{Discussion of the shape of the temporal afterpulse distribution}\label{sub:shape_disc}
As explained in section \ref{sec:AfterpulseProb} afterpulses are caused by ionization processes between the photocathode and the first dynode.
The time distribution of the afterpulses is dominated by the rather slow movement of the heavy ions with respect to the transit time of the photoelectrons.
Due to the nonuniform electric field of a hemispherical photomultiplier the travel time between the ionization location and the photocathode is not
sensitive to the ionization location \cite{ap_paper_coates}. The travel time can be calculated to
\begin{equation}
\label{eq:APtime}
t \propto \sqrt{ \frac{m}{\ell \cdot V_0}} \, .
\end{equation}
Here, $m$ and $\ell$ are the mass and charge of the ion and $V_0$ is the voltage between photocathode and first dynode.\\
Both temporal afterpulse distributions $\mathcal{E}_i$ and $\mu_i$ show the same peak structure (cf. figure \ref{fig_time_dist_bm}) with at least five
peaks with mean times approximately at \SI{0.52}{\upmu s}, \SI{2.1}{\upmu s}, \SI{4.9}{\upmu s}, \SI{6.3}{\upmu s} and \SI{7.9}{\upmu s}. 
This leads to the assumption that the afterpulses are mainly caused by five different types of ions.
According to calculations from Hamamatsu the peak at \SI{2.1}{\upmu s} is caused by Methane (${\rm CH_4}$) and the peak structure between
\SI{6}{}-\SI{8}{\upmu s} is caused by Caesium (Cs). Other possible candidates are Hydrogen (H) and Helium (He) as well as the cathode materials
Potassium (K) and Antimony (Sb). However, it's not possible to clearly assign these elements to the observed peaks.

%% file: sections/section_6.tex
\section{Conclusion}
We presented afterpulse timing measurements for 473 photomultipliers which served as one of the qualification tests for the
Double Chooz experiment. 
The selection specification  of the apfterpulse probability has been limited to \SI{10}{\%}  for each PMT. It was shown that all PMTs fulfilled
this requirement (figure \ref{fig_hist_AP_prob}).\\
The temporal distribution of the afterpulse occurence was determined for the number of expected afterpulses.
In a second analysis using the same data set the probability of measuring at least one afterpulse and the timing spectrum for the
expected number of first order afterpulses was computed.
Both methods' results showed as expected a strong correlation (cf. figure \ref{fig_total_ap_bm}) and were used to derive the
average charge transferred by an afterpulse $\overline{\ell}$.\\
More detailed timing and charge measurements of afterpulses were performed for a few PMTs using a different setup. 
The results of one PMT was presented, which represents well the general behaviour (figure \ref{f_hd3}).
Comparing the afterpulse timing spectra, both measurements show a good agreement for times greater than \SI{4}{\upmu s}. 
The additional exponential contribution in the second measurement has been identified to be caused by afterglowing effects of the LED.\\
The two different analysis methods used for the first measurement allowed us to determine the average number of photoelectrons contained by
afterpulses of $\overline{\ell} = \SI{(4.95 \pm 0.13)}{PE}$. As the charge measurements contain also LED afterglowing events, a direct comparison
of the average number of PE carried by afterpulses between the first and the second measurement was not possible.
However, the charge measurements indicate that afterpulses can carry charges in the range of 1 and 40 PE (figure \ref{f_hd3a} and \ref{f_hd3b}).\\
The measurements performed with different high voltages does not show a significant effect on the afterpulse time distribution shape.
Only a small linear correlation between the applied high voltage and the afterpulse probability can be observed.

%% file: sections/acknowledgments.tex
\section*{Acknowledgments}

This work is supported by the DFG (Deutsche Forschungsgemeinschaft).\\
We thank the whole Double Chooz PMT group for the excellent cooperation
and would like to point out particularly the important contributions
during the PMT calibration phase
of the CIEMAT group (Madrid, Spain) and the Double Chooz Japan group
(Hiroshima Institute of Technology, Kobe University, Niigata
University, Tokyo Institute of Technology, Tokyo Metropolitan
University, Tohoku Gakuin University, Tohoku University).